\documentclass[twocolumn,pra,aps,showpacs]{revtex4-1}
\usepackage{amssymb}
\usepackage{amsmath}
\usepackage{graphicx}

\begin{document}

\author{Lin Zhang}
\affiliation{College of physics and Information Technology, Shaanxi Normal University,
Xi'an 710061, China}

\title{Opto-mechanical measurement of micro-trap via nonlinear cavity enhanced Raman scattering spectrum}

\begin{abstract}
High-gain resonant nonlinear Raman scattering on trapped cold atoms within a
high-fineness ring optical cavity is simply explained under a nonlinear
opto-mechanical mechanism, and a proposal using it to
detect frequency of micro-trap on atom chip is presented. The
enhancement of scattering spectrum is due to a coherent Raman
conversion between two different cavity modes mediated by collective
vibrations of atoms through nonlinear
opto-mechanical couplings. The physical conditions of this
technique are roughly estimated on Rubidium atoms, and a simple quantum
analysis as well as a multi-body semiclassical simulation on this nonlinear Raman process
is conducted.
\end{abstract}

\pacs{42.50.Nn; 37.30.+i; 37.10.Vz; 42.65.Dr;}
\maketitle

\section{Introduction}
\label{sec1}%
Similar to the technology of integrated circuits on electronic
chips to control electrons, the micro-fabricated chips to manipulate
cold atoms and molecules \cite{Ron,Meek,Colombe}, ions and plasmas \cite%
{Stick,Shanhui} or massive nano-mechanical resonators \cite{Anetsberger} are
now being actively pursued towards precise quantum measurements \cite%
{Cronin,Riedel} and information processing \cite{Dumke,Birkl}. Combined with
different micro-fabricated elements \cite{Zoller}, atom chips provide
versatile integrated devices to precisely control particle's motion \cite%
{Folman}, biological cells \cite{Chiou}, chemical reactions or material
assembling \cite{Bonn,deMello,Jensen} on a specially fabricated substrate to
nanometer scales. Among these trends, controlling atomic motions in
near-field potential plays an important role in coherent manipulation of
atomic states. On atom chips, atoms can be confined above a surface where
the electric, magnetic or optical fields~\cite{Reichel} are built towards
coherent manipulation on both internal and external degrees of freedom. The
local forces produced on the chip can pack atoms into a micrometer-sized
region ($<1 \mu m$) to a condition that the internal transitions and the
external motions are intensely entangled \cite{Riedel,Colombe}. Indeed the
motion of the atoms under a quantum confinement is always the main focus in
all the above applications on atom chips.

Therefore, precise estimation of local potentials in which the atoms are confined
becomes very crucial to identify the motion of the atoms in small
scale. Considering modern optical trapping techniques \cite{Erikkson}, atoms
can freely be controlled by highly focused near electromagnetic fields~\cite%
{Hammes} on atom chips within microcavities and a real-time detection of local field along
certain dimensions is an essential aspect to estimate the spatial
distribution and the dynamic stability of confined atomic clouds under controls \cite%
{wieman1}.

In order to estimate the trapping parameters of micro-traps, many different
methods have been proposed, such as dynamical methods on atomic oscillations~%
\cite{Kim,Kim2,Moon}, the trap-loss rate method of atomic collisions \cite%
{Hoffmann}, the free-expansion method by temperature measurement \cite%
{Wallace} and optical spectroscopic methods via stimulated Raman
scattering (SRS) \cite{Grison,Brzozowski,Yoshikawa} or recoil induced
resonance on atomic velocity \cite{Meacher,Kozuma,Brzozowska}. Because
the field-induced micro-trap on atom chips is relatively flexible on small
scales, above mentioned mechanical methods on this system are difficult to
carry out in a high resolution limit. The conventional optical probing
methods are mainly based on stable absorptive spectrum related to
incoherent radiation of atoms limited by the saturation effect \cite%
{Razdan} and the gain of the probe mode for a detection is weak \cite{Kruse}%
. Although SRS is a well-established technique on condensed materials, it
has not been studied much in the context of atom-optics system \cite%
{Grison,Brzozowska,Louni}.

In this paper, we demonstrate a high-gain pump-probe Raman spectrum on a
trapped cold atomic ensemble \cite{Yang} and propose a more convenient and
sensitive method to diagnose micro-traps by using the coherent enhanced Raman
spectrum in high finesse microcavity. Due to atomic recoil motion, atomic
oscillations in the micro-trap will be actuated and closely entangled with
resonant cavity modes through scattering in atomic absorption-emission cycles~\cite%
{Meacher}. The dynamical backaction of the atomic oscillation dramatically
modifies the optical spectrum and its intensity is further enhanced
by a synchronization of the individual oscillators \cite{Cube} (equivalent
to building a collective mechanical mode). The fact that the optical spectrum of
trapped atoms in high-finesse cavity is enhanced by cooperative vibrations \cite{Hemmerich} is the problem
this paper seeks to address by developing a detection technique of microtrap
on atom chips.

\section{Model and mechanism}

\label{sec2} %
\begin{figure}[t]
\begin{center}
\includegraphics[width=0.45 \textwidth]{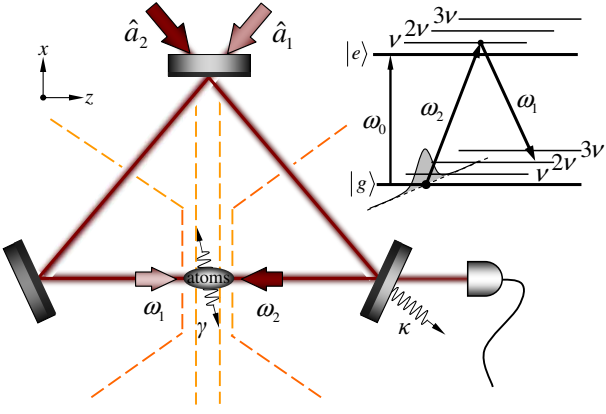}
\end{center}
\caption{The scheme of the probe-pump modes within a high-Q ring
cavity. The dashed line is the circuit used to generate microtraps on the
atom chip. The inset above right gives the level structure to demonstrate
stimulated Raman transition, where the wave packet stands for initial
thermal distribution and $\protect\nu$ is the trapping frequency.}
\label{Figure1}
\end{figure}

%%%
The model under consideration concerns about a cold atomic gas of $N$
identical atoms with transition frequency $\omega_0$, mass $m$ and
trapped by an external potential approximately treated as an one-dimensional
potential $V_{trap}=\frac{1}{2}m\nu_{z}^{2} z^{2}$, which corresponds to a
typical configuration on atom-chips~\cite{Su,Horak}. This is a reasonable assumption
on cold atomic gas (e.g. BEC) for that the thermal atoms will concentrate more on
lower vibration modes and the trap can be approximated by a
harmonic potential. The trapped atomic gas is enclosed in a high-Q
ring cavity towards employing a pump-probe scheme \cite{Himsworth}, in order
to diagnose the trapping frequency $\nu_z$ that characterizes the effective
trapping potential $V_{trap}$, by using two counter-propagating optical
fields, a strong pump mode with frequency $\omega_2$ to stimulate and a
weak probe mode with frequency $\omega_1$ to detect the atoms that are
trapped in a near-field potential as shown in Fig.\ref{Figure1}. %%%
%%%%%%%%%%%%%%%%%%%%%%%%%%
The direction of propagation for probe and pump modes are taken to be along $%
\pm{z}$ respectively, giving rise to the scheme of stimulated Raman
scattering process. Our objective is to see the dependence of
spectral gain of probe mode on trapping frequency $\nu_z$ that can be estimated by a
measurable quantity, namely the detuning between pump and probe
modes, $\delta_{21}=\omega_2-\omega_1$ and analyze the mechanical environment of the atomic motion.

This kind of pump-probe scheme within a strong atom-cavity coupling regime
in high-Q optical cavities is routinely considered in cavity QED experiments
and within the reach of practical realizability now~\cite%
{Kimble,Hemmerich,cavity}. For our current purpose the high-Q cavity
provides a low noise detection, which not only enhances atom-cavity coupling
but also increases the sensitivity of measurement by a factor of $1/(1 -R)$,
where $R$ is the reflectivity of cavity mirror~\cite{Ezeki}. Further, these
requirements of the system enable a usage of Bonifacio's collective atomic
recoil laser (CARL) model \cite{Boni1} to introduce an extra trapping
potential~\cite{Yang} for the micro-trap on atom chip. Describing
the two-level atomic gas with the Pauli spin operator $\hat{\sigma}^{0}$ and
their internal transitions by $\hat{\sigma}^{\pm}$, pump and probe modes by $%
\hat{a}_2,\hat{a}_1$, the system Hamiltonian is~\cite{Yang},
\begin{eqnarray}
\hat{H} &=&\sum_{j=1}^{N}\left( \frac{\hat{p}_{j}^{2}}{2m}+\frac{1}{2}m\nu
_{z}^{2}\hat{z}_{j}^{2}\right) +\hbar \omega _{0}\sum_{j=1}^{N}\hat{\sigma}%
^{0} _{j}+\hbar \omega _{1}\hat{a}_{1}^{\dagger }\hat{a}_{1}  \notag \\
&&+\hbar \omega _{2}\hat{a}_{2}^{\dagger }\hat{a}_{2} +i\hbar \left[ g_{1}%
\hat{a}_{1}^{\dagger}\sum_{j=1}^{N}\hat{\sigma} _{j}^{-}e^{-ik_{1}\hat{z}%
_{j}}-H.c.\right]  \notag \\
&&+i\hbar \left[g_{2}\hat{a}_{2}^{\dagger}\sum_{j=1}^{N}\hat{\sigma}%
_{j}^{-}e^{+ik_{2}\hat{z}_{j}}-H.c. \right],  \label{CARL}
\end{eqnarray}
where $g_{1} (g_2)$ is atom-probe(pump) coupling rate.

Before we embark upon estimating the potential, we shall now detail the
mechanism of high-gain Raman scattering spectrum and how it can lead to
estimating the trapping potential. The underlying mechanism of this
model depends on a nonlinear effect due to a
nonlinear coupling of atomic internal transitions with the vibration modes
of the micro-trap. Identifying the position and momentum as
\begin{equation*}
\hat{z}_{j}=\sqrt{\frac{\hbar }{2m\nu_z }}(\hat{b}_{j}+\hat{b}_{j}^{\dagger
}),\quad \hat{p}_{j}=-i\sqrt{\frac{\hbar m\nu_z }{2}}(\hat{b}_{j}-\hat{b}
_{j}^{\dagger }),
\end{equation*}%
with $\hat{b}_j$($\hat{b}_j^\dagger$) being phonon annihilation(creation)
operator of $j^{th}$ atom and obeys bosonic commutation relation. Moving to
the interaction picture with respect to the free part of Hamiltonian $\hat{H}%
_{0}=\sum_{j=1}^{N}\hbar \nu_z \hat{b}_{j}^{\dagger } \hat{b}_{j}+\hbar
\omega _{0}\sum_{j=1}^{N}\hat{\sigma}^{0}_{j}+\hbar \omega _{1}\hat{a}%
_{1}^{\dagger }\hat{a}_{1}+\hbar \omega _{2}\hat{a}_{2}^{\dagger } \hat{a}%
_{2}$, and using expanded Glauber displacement operator \cite{Vogel}
\begin{equation*}
\hat{D}(-i\eta )\equiv e^{-i\eta (\hat{b}_{j}^{\dagger }+\hat{b}
_{j})}=e^{-\eta ^{2}/2}\sum_{n,m}\frac{(-i\eta )^{n+m}}{n!m!}\hat{b}
_{j}^{\dagger n}\hat{b}_{j}^{m},
\end{equation*}
above Hamiltonian can be simplified to
\begin{eqnarray}
\hat{H}_{I} &=&i\hbar g_{1}e^{-\eta
_{1}^{2}/2}\sum_{j=1}^{N}\sum_{n,m}^{\infty }\frac{(-i\eta _{1})^{n+m}}{n!m!}
\hat{a}_{1}^{\dagger }\hat{b}_{j}^{\dagger n}\hat{b}_{j}^{m}\hat{\sigma}%
_{j}^{-}  \notag \\
&&\times e^{i(\omega _{1}+n\nu_z -m\nu_z -\omega _{0})t}-H.c  \notag \\
&&+i\hbar g_{2}e^{-\eta _{2}^{2}/2}\sum_{j=1}^{N}\sum_{n,m}^{\infty }\frac{
(i\eta _{2})^{n+m}}{n!m!}\hat{a}_{2}^{\dagger }\hat{b}_{j}^{\dagger n}\hat{b}
_{j}^{m}\hat{\sigma}_{j}^{-}  \notag \\
&&\times e^{i(\omega _{2}+n\nu_z -m\nu_z -\omega_{0})t}-H.c.  \label{HI}
\end{eqnarray}
Here the Lamb-Dicke parameter $\eta _{i}=k_{i}\sqrt{\hbar/%
2m\nu_z } \equiv\sqrt{\omega _{ri}/\nu_z }$ $(i=1,2)$ will further scale the
opto-mechanical coupling with $\omega_{ri}=\hbar k_i^2/2 m$ being the
single-photon recoil frequency. This interaction Hamiltonian makes it easier
to see the stimulated scattering initiated by cavity modes. Further the
scattering amplitudes oscillate at an effective frequencies as indicated in
the exponentials and depend on $\omega_{1,2}$. Since these frequencies can
be arbitrarily controlled, one can immediately establish a resonance
condition such that the time-dependence in the above Hamiltonian can be
coherently suppressed. Therefore we get,
\begin{equation*}
\omega _{1}-\omega _{0}+\left( n-m\right) \nu_z =0, \omega _{2}-\omega
_{0}+\left( n^{\prime }-m^{\prime }\right) \nu_z =0,
\end{equation*}
which precisely leads to a resonant condition
\begin{equation}
\delta _{21}=(n^{\prime }-m^{\prime }+m-n)\nu_z =l\nu_z ,\quad l=0,\pm 1,\pm
2,\cdots,  \label{Ra}
\end{equation}
where the integers $l$ are the resonant scattering orders that can be used
to label the gain peaks. We call this relation Raman resonant condition
because the relation comes from Raman transitions between different cavity modes
mediated by vibration modes of the trap ($\hat{b}$) and the gain shares a similar mechanism with
Raman spectroscopy. When this condition is satisfied there occurs a maxima in
the output spectrum, introduced in next section, and can be readily
measured in a real experiment.

In addition, assuming $k_{1}\approx k_{2}=k$ leading to $\eta _{1}\approx
\eta _{2}=\eta $ and $g_{1}\approx g_{2}=g$ for simplicity, which we do for
numerical simulations, in resonant condition the Hamiltonian can be further
simplified to
\begin{eqnarray}
\hat{H}_{I} &=&i\hbar ge^{-\eta ^{2}/2}\Bigg[\hat{a}_{1}^{\dagger
}\sum_{j=1}^{N}\hat{\sigma}_{j}^{-}\sum_{n,m}^{\infty }\frac{(-i\eta )^{n+m}%
}{n!m!}\hat{b}_{j}^{\dagger n}\hat{b}_{j}^{m}  \notag \\
&&+\hat{a}_{2}^{\dagger }\sum_{j=1}^{N}\hat{\sigma}_{j}^{-}\sum_{n,m}^{%
\infty }\frac{(i\eta )^{n+m}}{n!m!}\hat{b}_{j}^{\dagger n}\hat{b}_{j}^{m}%
\Bigg]-H.c..  \label{Hrwa}
\end{eqnarray}%
In the high-Q cavity with a strong atom-photon coupling strength \cite%
{Kimble}, above Hamiltonian demonstrates an intense entanglement of the
vibration mode $\hat{b}_{j}$ with internal transition $\hat{\sigma}_{j}^{\pm
}$ through a nonlinear mechanical coupling coefficients $\eta ^{n}$. If $%
\eta \ll 1$, only lower vibration modes are involved and the Hamiltonian can
be expanded by a power series of $\eta $ where the high-order Raman
transitions are significantly depressed. In this case Eq.(\ref{Hrwa}) will
be simplified and a further analysis on the intensity of the peaks due to
coherent transitions between lower vibration modes is possible within this
$N$-body system and only the atoms distributed on lower vibration modes
are critical in determining the gain of the corresponding Raman peak.
However, height estimation of the peak $l$ in the Raman spectrum becomes
difficult when the Lamb-Dicke parameter $\eta \sim 1$ leading to a
photon recoil frequency comparable to trap frequency $\nu _{z}$ and
Raman transitions between higher vibration levels should be included.
Nevertheless the simplified Hamiltonian in
Eq.(\ref{Hrwa}) gives an insight into the Raman process occurring due to coherent
atomic motion.

A specific case which can highlight the elusive opto-mechanical coupling between the cavity mode
and the atomic vibration is under the conditions of
far off-resonant with low atomic saturation when the upper
atomic levels can be adiabatically eliminated. Neglecting the weak probe and treating
all the atoms in a homogenous pumping field, the
far-off intense pumping mode induces an effective atomic polarization \cite%
{Domokos}%
\begin{equation*}
\hat{\sigma}_{j}^{-}\approx \frac{g}{i\delta _{20}-\gamma }\hat{a}_{2},
\end{equation*}%
which leads to a Hamiltonian
\begin{equation}
\label{Hopt}
\hat{H}_{I}\approx \hbar \left( U_{0}-i\Gamma _{0}\right) \hat{a}%
_{2}^{\dagger }\hat{a}_{2}\sum_{j=1}^{N}e^{i\eta (\hat{b}_{j}^{\dagger }+%
\hat{b}_{j})}+H.c.,
\end{equation}%
where $\delta_{20}=\omega_2-\omega_0$ is the detuning of the atom
from the pumping field and $\gamma$ is the atomic
spontaneous emission rate. The above Hamiltonian
clearly describes an opto-mechanical interaction that is extensively studied on cold atoms.
Eq.(\ref{Hopt}) demonstrates that an opto-mechanical model for cold atoms
are actually related to coherent motion of $N$ atoms which is evaluated by the
collective order parameter $R=\sum_{j=1}^{N}e^{ik\hat{z}_{j}}$.
With large-detuned condition of $\delta_{20}\gg \gamma$, the light-shift
coefficient $U_0$ and the incoherent scattering rate $%
\Gamma_0 $  reduce to%
\begin{subequations}
\label{polarization}
\begin{eqnarray}
U_0 &=&\frac{g^{2}\delta _{20}}{\delta _{20}^{2}+\gamma ^{2}}\approx \frac{%
g^{2}}{\delta _{20}}, \\
\Gamma_0  &=&\frac{g^{2}\gamma }{\delta _{20}^{2}+\gamma ^{2}}\approx 0,
\end{eqnarray}%
\end{subequations}
respectively. Therefore an effective Hamiltonian under large-detuned pumping field reads
\begin{equation*}
\hat{H}_{I}\approx \hbar \frac{2g^{2}}{\delta _{20}}\hat{a}_{2}^{\dagger }%
\hat{a}_{2}\sum_{j=1}^{N}\cos \left( k\hat{z}_{j}\right),
\end{equation*}
with an effective opto-mechanical coupling clearly depending on the collective vibration modes
of the $N$ atoms. If all the atoms are synchronized on a same vibration mode,
an extreme linear coupling rate is obtained by
\begin{equation}
\label{G0}
G=\frac{2g^{2}N}{\delta _{20}},
\end{equation}
whose strength can be greatly enhanced by the population numbers, $N$, on the vibration mode.
Since the above interaction Hamiltonian is not limited with only the linear in $\hat{z}_j$,
one can in general employ the methods in this work for high-order
nonlinear displacements (Quadratic in position, for example \cite{Milburn}) of a
mechanical oscillator due to radiation pressure of pumping
mode proportional to intensity $\hat{a}^{\dag}_2\hat{a}_2$.

\section{Gain spectrum and detection}

We now proceed with a further analysis by numerically simulations on the dynamical
behavior of this multiple-body system on a general basis. The evolution of the density
matrix $\hat{\rho}$ with cavity damping $\kappa _{1}(\kappa_{2})$ and atomic
spontaneous emission rate $\gamma$ can be described with a master equation
\begin{equation}
\frac{d\hat{\rho}}{dt}=\frac{1}{i\hbar }[\hat{H},\hat{\rho}]+\hat{\mathcal{L}
}_{j}\hat{\rho}+\hat{\mathcal{L}}_{f}\hat{\rho},  \label{master}
\end{equation}%
where the Louivillians
\begin{eqnarray}
\hat{\mathcal{L}}_{j}\hat{\rho} &=&\frac{\gamma}{2}\sum_{j=1}^{N}\left( 2%
\hat{\sigma}_{j}^{-}\hat{\rho}\hat{\sigma}_{j}^{+} -\hat{\sigma}_{j}^{+}
\hat{\sigma}_{j}^{-}\hat{\rho}-\hat{\rho} \hat{\sigma}_{j}^{+}\hat{\sigma}%
_{j}^{-}\right) ,  \notag \\
\hat{\mathcal{L}}_{f}\hat{\rho}&=&\frac{1}{2}\sum_{i=1,2}\kappa _{i}\left( 2%
\hat{a}_{i}\hat{\rho}\hat{a}_{i}^{\dagger }-\hat{a}_{i}^{\dagger }\hat{a}%
_{i} \hat{\rho}-\hat{\rho}\hat{a}_{i}^{\dagger }\hat{a}_{i}\right) ,  \notag
\end{eqnarray}%
and the semiclassical behavior can be investigated by the dynamic motion of expectation values $%
\langle \dot{O}\rangle=Tr(\dot{\hat{\rho}} \hat{O})$. We use the full
Hamiltonian Eq.~(\ref{CARL}) so as to include higher order transitions as
explained towards the end of previous section. Thus the equations of motion
for relevant observables are,
\begin{subequations}
\label{eq2}
\begin{eqnarray}
\frac{d\theta _{j}}{d\tau } &=&p_{j} , \\
\frac{dp_{j}}{d\tau } &=&-\nu ^{2}\theta _{j}-A_{1}^{\ast }\sigma
_{j}e^{-i\theta _{j}}-A_{1}\sigma _{j}^{\ast }e^{i\theta _{j}}  \notag \\
&&+A_{2}(\sigma _{j}+\sigma _{j}^{\ast }) , \\
\frac{d\sigma _{zj}}{d\tau } &=&2\rho \left[ \left( A_{1}^{\ast }e^{-i\theta
_{j}}+A_{2}\right) \sigma _{j}+\sigma _{j}^{\ast }\left( A_{1}e^{i\theta
_{j}}+A_{2}\right) \right]  \notag \\
& &-\Gamma \left( \sigma _{0j}-1\right) \\
\frac{d\sigma _{j}}{d\tau } &=&i\left( \Delta _{20}+\frac{p_{j}}{2}\right)
\sigma _{j}-\rho \sigma _{0j}\left( A_{1}e^{i\theta _{j}}+A_{2}\right)
\notag \\
&& -\Gamma \sigma _{j} , \\
\frac{dA_{1}}{d\tau } &=&i\Delta _{21}A_{1}+\frac{1}{N}\sum_{j=1}^{N}\sigma
_{j}e^{-i\theta _{j}}-\kappa A_1 ,
\end{eqnarray}%
\end{subequations}%
where the pump mode $A_2$ is in a strong coherent state and excluded
from Eq.(\ref{eq2}), sustaining by a stabilized locking driven field \cite%
{cavity}.  The dimensionless variables introduced in above equations are $%
\theta_{j}=2k\left\langle \hat{z}_{j}\right\rangle $, $p_{j}= \frac{
\left\langle\hat{p}_{j}\right\rangle }{\hbar k\rho }$, $\sigma
_{j}=\left\langle \hat{\sigma }_{j}^{-}e^{ik_{2} \hat{z} _{j}}\right\rangle
e^{i\omega _{2}t}$, $\sigma _{0j}=-2\left\langle \hat{ \sigma}%
^0_{j}\right\rangle $, $A_{1}=\frac{\left\langle \hat{a}_{1}\right\rangle }{%
\sqrt{N\rho }}e^{i\omega_{2}t}$, $\Delta_{21}=(\omega _{2}-\omega
_{1})/(\omega _{r}\rho )$, $\Delta _{20}=(\omega _{2}-\omega _{0})/(\omega
_{r}\rho )$, $\nu=\nu_z/(\omega_r \rho)$, $\Gamma=\gamma/(\omega_r \rho)$, $%
\kappa= \kappa_1/(\omega_r \rho)$ and $\tau =\omega _{r}\rho t$ assuming $%
k_{1}\approx k_{2}=k$ and $g_{1}\approx g_{2}=g$, for simplicity. The
scaling parameters $\omega _{r}$ and $\rho $ are photon recoil frequency
$\omega _{r}=(2\hbar k)^{2}/2m$ and collective coupling constant $\rho =( g%
\sqrt{N}/\omega _{r}) ^{2/3}$ respectively \cite{Boni1}. Among the above
rescaled parameters, $\Delta _{20}$, the frequency detuning of pump mode
from the atomic transition, and $\Delta _{21}$, the detuning between the
pump mode and the probe mode, can be modulated in the experiment by
frequency chirping method.
\begin{figure*}[tbp]
\begin{center}
\includegraphics[width=165pt]{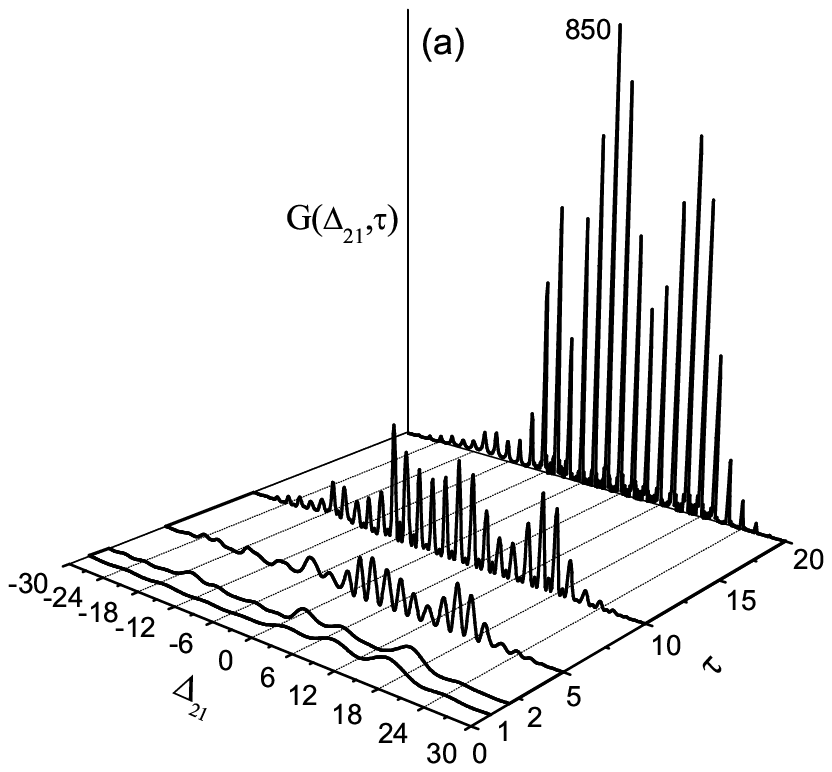} %
\includegraphics[width=155pt]{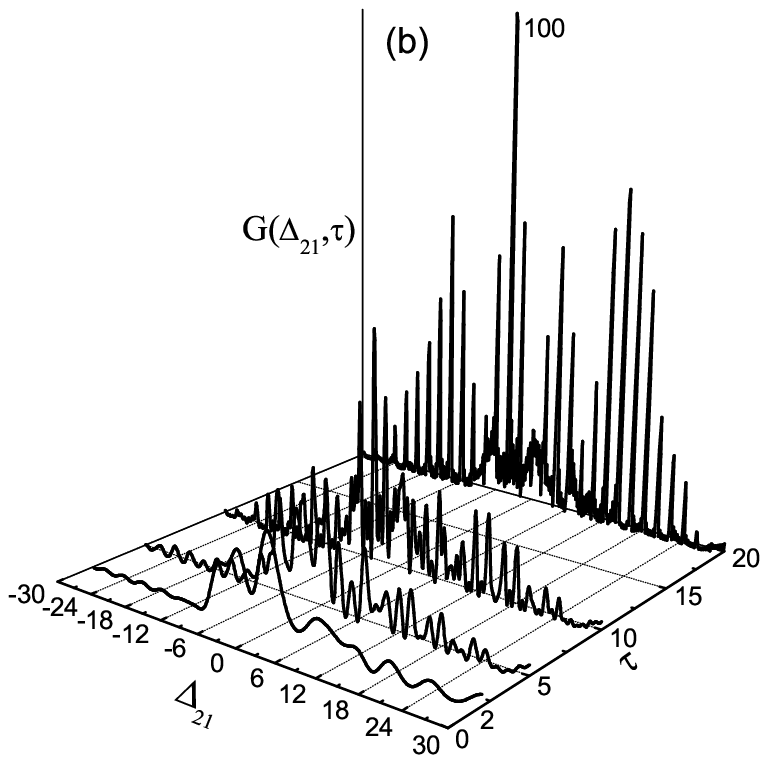} %
\includegraphics[width=155pt]{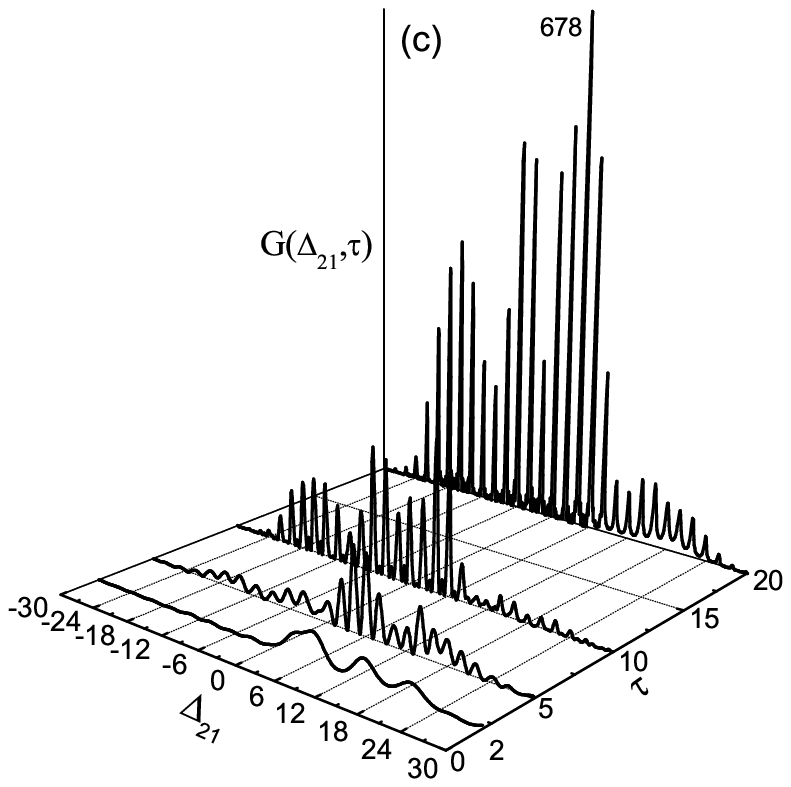}
\end{center}
\caption{The simulation of gain spectrum of the probe mode at different
pumping time $\protect\tau$ in a harmonic trap with the trapping frequency $%
\protect\nu=2$. (a) Red-detuned driving with $\Delta_{20}=-15$;
(b) Resonant case with $\Delta_{20}=0$; (c) Blue-detuned driving
with $\Delta_{20}=15$. The other parameters are $\protect\rho=3$, $%
A=2$, $\Gamma=1$, $\protect\kappa=0.01$. The number of the atoms in the trap
is only $N=100 $. All the parameters are scaled by the collective recoil
frequency $\protect\rho \protect\omega_r$.} \label{Figure2}
\end{figure*}
The figure of merit, as in conventional CARL theory~\cite{Boni1}, is the time dependent
gain function of the probe field $A_1$ and is given as
\begin{equation}
G\left( \Delta _{21},\tau \right) =\frac{\left\vert A_{1}\left( \tau \right)
\right\vert ^{2}-\left\vert A_{1}\left( 0\right) \right\vert ^{2}}{
\left\vert A_{1}\left( 0\right) \right\vert ^{2}},
\end{equation}
where $A_{1}\left( 0\right) $ is the initial amplitude of probe mode.

For our simulation, we use Rubidium atoms and the Lamb-Dicke parameter $%
\eta= \sqrt{\omega_{r}/\nu} \approx 0.4$ with maximum resonant number
excited in Eq.(\ref{Ra}) is about $l\approx\pm 15$ as shown in Fig.\ref%
{Figure2}. Therefore the contributions of the high-order Raman transitions
will not be omitted beyond the Lamb-Dicke approximation and the use of
original Hamiltonian Eq.~(\ref{CARL}) in equations of motion for
observables Eq.~(\ref{eq2}) is needed. It is worth noting that, if the
density of the atomic gas is increased in a stronger trap, the Lamb-Dicke
parameter $\eta$ will be small enough that the Raman transition will be
closely concentrated around the central region of lower scattering numbers.

Our calculations on this system as shown in Fig.\ref{Figure2} demonstrates
a high gain of the weak probe mode \cite{Yang,Zhang}, and reveals a cavity
enhanced Raman spectrum with a regular pectinate profile. The high-gain
peaks of the spectrum are closely related to the collective atomic motion in
the trap with a gain magnitude enhanced by population occupations of the
atoms on different vibration modes \cite{Jessen}. Due to atomic recoil,
the thermal atoms with an initial Gaussian velocity distribution (which
is randomly distributed to $N$ atoms at the initial time, see inset of Fig.\ref%
{Figure1}) are quickly redistributed to different oscillation modes through
optical scattering supported by the high-Q cavity. Then the probe
mode is amplified gradually by coherent Raman transitions between
different vibration modes of the atoms and will be further improved by a
synchronization of collective motion (see Eq.(\ref{G0})). After the
depletion of thermal atoms, line width of the gain peak is reduced by an
elimination of Doppler broadening (see Fig.\ref{Figure2}) \cite{Letokhov}
and replaced by a comb like profile. Although this gain enhancement of the
atoms in the trap has been observed in many experiments \cite%
{Kruse,Hemmerich,Slama,Ringot}, the Raman spectrum and the detection technique under this scheme is
not purposely explored. In the present case, the enhanced gain of the probe
mode detected at different times after a continuous-wave pumping turned on
can be obtained by a frequency scanning process with chirping probe laser(see
Ref.\cite{cavity}). Fig.\ref{Figure2}(a) is the scanning spectrum of the
probe mode obtained under the condition that the pump is red-detuned from the
atomic transition frequency, while Fig.\ref{Figure2}(b) is in a resonant case
and Fig.\ref{Figure2}(c) is blue detuned.

Although all the detunings in Fig.\ref{Figure2} gives high resolution
spectra with sharp peaks, the gain magnitude of the probe mode are actually
in a different order after a same pumping
time, $\tau$. For resonant pumping, the maximum height of the peak at time $%
\tau=20$ is about $10^2$, which is almost one order smaller than that in the
off-detuned spectrum (the gain magnitudes of the highest peaks are labeled
in Fig.\ref{Figure2}). This difference is because that an off-detuned pumping
will increase the absorption-emission recycling frequency to an effective
Rabi frequency $\Omega_R\approx\sqrt{\Delta_{20}^{2}+(2\rho A_2)^2+\Gamma^2}$
and improve the saturation magnitude of atomic dispersion, indicating an
optimal spectrum for dispersive atomic medium (see Eq.(\ref{polarization}))
to establish coherent vibration modes.
The off-detuned pumping also can
suppress the spontaneous emission which introduces random motions into
the coherent atomic oscillation in the trap and, thus, diminishes the
spectral magnitude with a lower resolution. The other difference of these
spectra is that the Raman transition shifts more into Stokes region by the
red-detuned driving (Fig\ref{Figure2}(a)) and shifts more into
anti-Stokes region under the blue-detuned pumping (Fig\ref{Figure2}(c)).
This is clearly due to the cooling and the heating effects of the
red-detuned and the blue-detuned driving on atomic vibration modes,
respectively. %%%%%%%%%%

The comb-like structure under all of the three pumping conditions implies a
robust detection method of trapping potentials. By analyzing the frequency
difference between two resonant peaks with a spectrometer, the intensity of
a micro-trap along the pump-probe direction can be easily obtained. Fig.\ref%
{Figure2} demonstrates that the minimum unit distance between two
neighboring peaks of the gain spectrum is $2$, which is exactly the
trapping frequency $\nu$ along pumping direction. Clearly, the positions of the
peaks in Fig.\ref{Figure2} indicate a resonant condition of the pump-probe
detuning of
\begin{equation}
\Delta_{21}=l\nu, \quad l=0, \pm 1, \pm 2,\cdots,  \label{Raman}
\end{equation}
and which definitely verifies quantum analysis Eq.~(\ref{Ra}).

Fig.\ref{Figure2} also reveals that the gain of a probe mode is a time
developing spectrum after the pumping mode is switched on, increasing with
time first and, finally, stabilized by the saturation of synchronization and
the damping process of the probe field and the atoms. Our simulations
verify further that the response times to establish a high gain spectrum
are very quick for all of the three pumping conditions. Without
considering the isotope fraction and hyperfine structure, a rough response
time of this spectrum on Rubidium atoms can be estimated. If the density of
Rubidium gas loaded in the micro-trap is about $10^{2}$ atoms/cm$^{3}$ ($%
\rho =3$) and the atoms are driven by a strong field on $5^{2}P_{3/2}
\longleftrightarrow 5^{2}S_{1/2}$ transition with $\lambda =780nm$, then the
collective recoil frequency $\rho \omega_{r}$ of the gas is about $%
2.88\times 10^{5}$ Hz, and the time interval needed to develop a detectable
high-gain spectrum is about $69.4$ microseconds ($\tau =20$), and which is a
reasonable time in a typical experiment. If the density of the atom is
increased to $10^{8}$ atoms/cm$^{3}$ in a stronger micro-trap, the building
time for a high-gain spectrum will be only $1.0$ microseconds. Increasing
the atom-field coupling $g$ in the high-Q cavity will also speed up the
synchronization process of collective motion and shorten the response
time of the spectrum. Compared with other detection methods \cite%
{Moon,Hoffmann}, this proposal provides a quicker and precise measurement
on potential parameters for a micro-trap if the system is connected to a
computer assisted spectroanalysis with a chirping probe fields at an optimal
ramping rate~\cite{cavity}.

When the response time of the gain reduces below one nanosecond, a pulse
wave pump-probe detection might be employed on a highly dense atomic gas in
a strong trap \cite{Polli}. Surely this pump-probe Raman scattering method
can also be used to analyze other types of traps (e.g., anharmonic traps) as
long as the collective motion of the atoms can be coherently stimulated by an electric,
magnetic or optical force during the mode scanning process of the
detection.

\section{Conclusion}

\label{sec3}In order to identify the mechanical stiffness of the local micro-trap on
atom chips, an opto-mechanical measurement on the frequency
of the trap based on a high-gain Raman scattering spectrum of probe mode in a ring cavity has been proposed. This Raman
scattering spectrum owns a distinct pectinate profile with high resonant
peaks, narrow spectral lines and develops with the pumping time. The
spectral gain is derived from a resonant Raman scattering of pumping mode to probe mode on atomic
vibration motion and enhanced by synchronized collective motion
of atoms (corresponding to coherent vibration mode) established in the trap. The narrowed peak width comes from a
removal of the Doppler broadening through a coherent distribution of
initial thermal atoms into different vibration modes in the trap. This
Raman pump-probe spectroscopy for depth measurement of micro-traps is
quick and more sensitive than that of the traditional methods, and can be further
developed into a rapid detecting method on surface vibration dynamics or on
surface plasmon oscillations. Because trapping and guiding ions, atoms,
molecules or even plasma with nanofabricated circuits on atom chips are all
closely related to collective motions, the method we presented here
is universal to detect different micro-traps as long as the coherent motion
can be efficiently coupled with internal atomic transition by the
nonlinear Raman scattering process. A concise quantum analysis of this high-gain Raman
spectrum reveals a nonlinear opto-mechanical coupling of atomic vibration
with two cavity modes mediated by atomic internal level transitions. This nonlinear gain mechanism also provides
an insight into the stimulated Raman spectrum of vibrating atoms on
condensed materials or gives a clue on the enhanced Raman spectrum of bound
electrons on a structured metal surface.

Certainly, the distinct profile of the Raman scattering spectrum of probe mode in
this paper depends on how well one can satisfy the approximations we adopted
in this work. The first approximation is an assumption of a one-dimensional
harmonic potential for cold trapped atomic gases. A more general trapping
potential on atom chips, often taking a 3-D periodic arrangement, will lead
to a more complicated spectral structure with nonuniform peaks and more
broadened lines or even optical band. Nevertheless the proposed method can still provide useful
information about the intensity of the trapping potentials along a certain
direction because of its opto-mechanical mechanism. The second approximation is related to the two-level assumption
of the $N$ trapped identical atoms which omits also the isotope fraction
\cite{Razdan,Himsworth} and the hyperfine level structure \cite{Smith,Zigdon}
which could lead to different Raman resonant relations of Eq.(\ref{Ra}) and
results in an overlap of different groups of Raman peaks. Finally, the
spectral line broadening mechanisms by atomic collision, random spontaneous
emission, beam collimation or by source linewidth \cite{Himsworth} are all
neglected in the calculation of the Raman scattering gain spectrum.

\section*{Acknowledgement}

This work was supported by the Shaanxi Provincial Natural Science Foundation
SJ08A12, PRC.%%%%

\end{document}